# Boson/Fermion Janus Particles

Roumen Tsekov

Department of Physical Chemistry, University of Sofia, 1164 Sofia, Bulgaria

Thermodynamically, bosons and fermions differ by their statistics only. A general entropy functional is proposed by superposition of entropic terms, typical for different quantum gases. The statistical properties of the corresponding Janus particles are derived by variation of the weight of the boson/fermion fraction. It is shown that di-bosons and anti-fermions separate in gas and liquid phases, while three-phase equilibrium appears for poly-boson/fermion Janus particles.

Janus particles consist of distinct parts with different physical properties. Such particles, based on wettability, are very popular in the modern colloidal nanotechnologies [1]. The scope of the present paper is to explore properties of another class of Janus particles, made of bosons and fermions, bonded somehow each other. Since the latter are ideal quantum gases, there is no interaction in the traditional sense between particles via an interaction potential. The difference between bosons and fermions is statistical and, hence, it is governed by the relevant entropies. Traditionally, in the Boltzmann statistics the occupation number $n_k$ is derived via variation of the system entropy $S$ at constant energy $E = \sum n_k \varepsilon_k$ and number of particles $N = \sum n_k$. The well-known expressions for the entropy of quantum gasses [2] can be generalized by the following functional

$$S/k_B = -\sum n_k \ln n_k + \alpha_B \sum (1+n_k)\ln(1+n_k) - \alpha_F \sum (1-n_k)\ln(1-n_k) \qquad (1)$$

The first term here is the Boltzmann entropy, while the other two terms are typical additives for bosons and fermions, respectively. Two free parameters $\alpha_B$ and $\alpha_F$ are introduced in Eq. (1) to vary the boson/fermion fraction in the Janus particle. The standard conditional variation of Eq. (1) yields

$$-\ln n_k + \alpha_B \ln(1+n_k) + \alpha_F \ln(1-n_k) = \beta(\varepsilon_k - \mu) \qquad (2)$$

where the Lagrange multipliers, the reciprocal thermodynamic temperature $\beta$ and the chemical potential $\mu$, can be obtained from the conservation of energy and of number of particles in the system. Equation (2) can be easily rearranged to obtain

$$B_k = \frac{n_k}{(1+n_k)^{\alpha_B}(1-n_k)^{\alpha_F}} \qquad (3)$$

where $B_k \equiv \exp[\beta(\mu - \varepsilon_k)]$ is the Boltzmann factor. Solving Eq. (3) yields a family of occupation numbers $n_k(\alpha_B, \alpha_F)$ for different boson/fermion fractions in the Janus particles. In the classical case Eq. (3) reduces to the Maxwell-Boltzmann distribution $n_k(0,0) = B_k$, which always holds in the dilute case $n_k \ll 1$. Generally, the function (3) possesses extrema at two points

$$n_\pm = \frac{\alpha_B - \alpha_F \pm \sqrt{(\alpha_B - \alpha_F)^2 - 4(\alpha_B + \alpha_F - 1)}}{2(\alpha_B + \alpha_F - 1)} \qquad (4)$$

which can be reverted to $\alpha_B = (1 + n_+ + n_- + n_+ n_-)/2n_+ n_-$ and $\alpha_F = (1 - n_+ - n_- + n_+ n_-)/2n_+ n_-$.

Let us start first with pure bosons, where $\alpha_F \equiv 0$ and Eq. (3) reduces to

$$B_k = n_k/(1+n_k)^{\alpha_B} \qquad (5)$$

The standard bosons follow the Bose-Einstein distribution $n_k(1,0) = B_k/(1-B_k)$, which predicts condensation at $B_k = 1$. In the case of $\alpha_B < 1$ the occupation number increases monotonously by increase of the Boltzmann factor $B_k$ and these quantum gases resemble more or less the classical Boltzmann gas. Hence, particles as anti-bosons with $n_k(-1,0) = (\sqrt{1+4B_k} - 1)/2$ and half-bosons with $n_k(1/2,0) = B_k(\sqrt{4+B_k^2} + B_k)/2$ cannot condensate. As seen in Fig. 1 the dependence of the Boltzmann factor $B_k$ on the occupation number passes through maximum at $n_- = 1/(\alpha_B - 1)$ in the case of $\alpha_B > 1$. This indicates a separation in two phases (Bose gas and liquid), because there are two occupation numbers at $B_k$ below the maximum. For instance, in the case of di-bosons with $\alpha_B = 2$ the maximum appears at $n_- = 1$ and equals to $B_- = 1/4$. Such behavior is not surprising, however, since the entropy additive of bosons corresponds to entropic attraction, leading to separation in two phases of poly-boson gases with $\alpha_B > 1$.

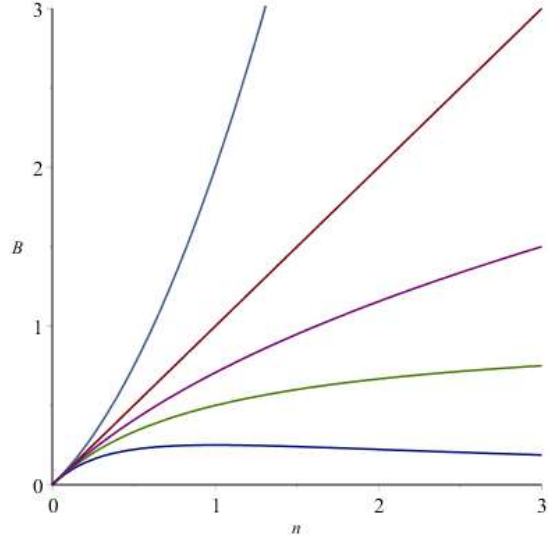

**Fig. 1** Plots of $n_k(-1,0)$ (light blue: anti-boson), $n_k(0,0)$ (brown: Boltzmann), $n_k(1/2,0)$ (purple: half-boson), $n_k(1,0)$ (green: standard boson) and $n_k(2,0)$ (dark blue: di-boson)

Let us consider now pure fermions, where $\alpha_B \equiv 0$ and Eq. (3) reduces to

$$B_k = n_k / (1 - n_k)^{\alpha_F} \tag{6}$$

The standard fermions obey the Fermi-Dirac distribution $n_k(0,1) = B_k / (1 + B_k)$, where the occupation number is always lower than 1, due to the exclusion principle. In the case of $\alpha_F > 0$ the fermionic character increases gradually by increase of $\alpha_F$. Thus, di-fermions with occupation number $n_k(0,2) = 1 - (\sqrt{1 + 4B_k} - 1)/2B_k$ repel stronger than half-fermions with occupation number $n_k(0,1/2) = B_k(\sqrt{4 + B_k^2} - B_k)/2$. As is seen from Fig. 2, the dependence of the Boltzmann factor $B_k$ on the occupation number passes through maximum at $n_- = 1/(1-\alpha_F)$ in the case of $\alpha_F < 0$. For instance, at $\alpha_F \equiv -1$ the maximum appears at $n_- = 1/2$ and equals to $B_- = 1/4$. Therefore, anti-fermions separate in two phases, similar to di-bosons. This is expected, since the entropy additive of fermions corresponds to entropic repulsion. Thus, anti-fermion gases can separate in two phases, since in this case the fermion entropy reverts to entropic attraction.

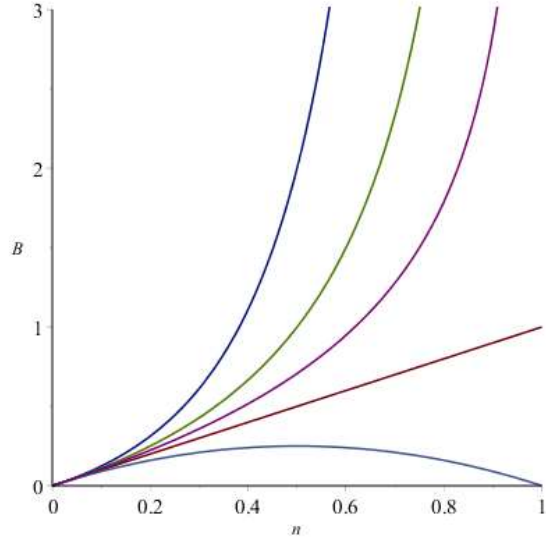

**Fig. 2** Plots of $n_k(0,2)$ (dark blue: di-fermion), $n_k(0,1)$ (green: standard fermion), $n_k(0,1/2)$ (purple: half-fermion), $n_k(0,0)$ (brown: Boltzmann) and $n_k(0,-1)$ (light blue: anti-fermion)

Let us consider now some boson/fermion Janus particles. Looking at Eq. (3) it easy to see that the fermion effect of restrictive, since in any case $n_k \leq 1$ for the Janus particles. For example, the distribution for half-boson/half-fermion Janus monomers is $n_k(1/2,1/2) = B_k / \sqrt{1+B_k^2}$. No surprises are expected in the occupation numbers of boson/fermion $n_k(1,1) = (\sqrt{1+4B_k^2} - 1)/2B_k$ and of anti-boson/fermion $n_k(-1,1) = (\sqrt{1+6B_k + B_k^2} - 1 - B_k)/2$ Janus dimers. As is seen from Fig. 3, the distributions for boson/anti-fermion and anti-boson/anti-fermion Janus dimers are also predictable. To find out some interesting examples one should consider the two possible solutions of Eq. (4) inside the range $0 < n_\pm < 1$. If one choses, for instance, $n_- = 1/3$ and $n_+ = 2/3$ it follows that they correspond to $\alpha_B = 5$ and $\alpha_F = 1/2$. Therefore, penta-boson/half-fermion Janus particles separates in three phases. Similar behavior shows tetra-boson/one-third-fermion Janus particles with $n_- = 1/2$ and $n_+ = 3/5$ or hexa-boson/fermion Janus particles with $n_- = 1/3$ and $n_+ = 1/2$. As is shown in Fig. 4, the middle composition corresponds to a Bose liquid, while the other two phases are diluter Bose and denser Fermi gases.

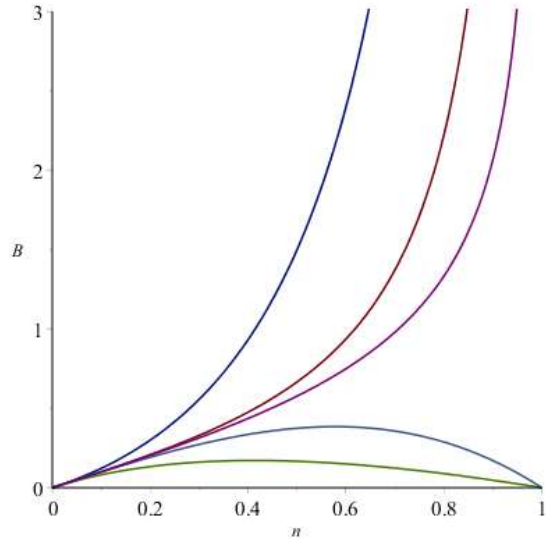

**Fig. 3** Plots of $n_k(-1,1)$ (dark blue: anti-boson/fermion), $n_k(1,1)$ (brown: boson/fermion), $n_k(1/2,1/2)$ (purple: half-boson/half-fermion), $n_k(-1,-1)$ (light blue: anti-boson/anti-fermion) and $n_k(1,-1)$ (green: boson/anti-fermion)

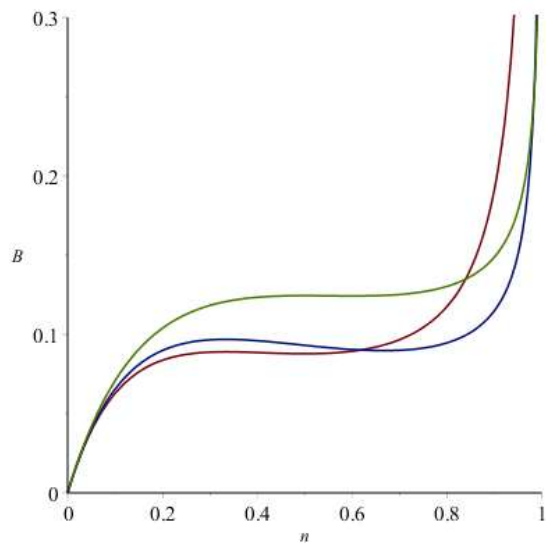

**Fig. 4** Plots of $n_k(4,1/3)$ (green: tetra-boson/one-third-fermion), $n_k(5,1/2)$ (blue: penta-boson/half-fermion) and $n_k(6,1)$ (brown: hexa-boson/fermion)

It is also of interesting to describe the dynamics of the Janus particles distribution. Not far away from equilibrium the chemical potential can be written in the form of Eq. (2)

$$\mu = \varepsilon_p + k_B T[\ln n_p - \alpha_B \ln(1+n_p) - \alpha_F \ln(1-n_p)] \qquad (7)$$

The Einstein relativity expression for the particle kinetic energy $\varepsilon_p = Mc^2$ is considered for generality, where $M = \sqrt{m^2 + p^2/c^2}$ is the relativistic mass of the particle with momentum $p$. Due to conservation of the number of particles, the non-equilibrium occupation number $n_p(t)$ of the Janus particles satisfies compulsory the continuity equation $\partial_t n_p = -\partial_p j_p$ in the momentum space. In the frames of non-equilibrium thermodynamics, the generalized flow can be presented in the form $j_p/n_p = -\gamma \partial_p \mu$, where the positive friction coefficient $\gamma$ can depend on $n_p$ as well. Substituting this expression, completed by Eq. (7), into the continuity equation yields

$$\partial_t n_p = \partial_p (n_p \gamma p / M + D_p \partial_p n_p) \qquad (8)$$

This is a relativistic Fokker-Planck equation, where an effective diffusion coefficient in the momentum space is introduced

$$D_p \equiv \gamma k_B T[1 - \alpha_B n_p / (1+n_p) + \alpha_F n_p / (1-n_p)] \qquad (9)$$

It is evident that the boson and anti-fermion 'attraction' reduces $D_p$, while fermion and anti-boson 'repulsion' increases the effective diffusion coefficient. Note that $D_p$ can be negative at some occupation numbers, how it is shown in Fig. 5. Since $D_p(n_\pm) = 0$, there is probably spinodal decomposition in the range $n_- < n_p < n_+$. However, in any case the equilibrium solution of Eq. (8) is a Jüttner-like distribution [3]. The non-linear Eq. (8) can be linearized by employing the equilibrium expression for the occupation number in the friction and effective diffusion coefficients, e.g. for bosons $D_p = \gamma k_B T / (1+n_p) \simeq \gamma k_B T (1-B_p)$, fermions $D_p = \gamma k_B T / (1-n_p) \simeq \gamma k_B T (1+B_p)$ and half-boson/half-fermion Janus particles $D_p = \gamma k_B T / (1-n_p^2) \simeq \gamma k_B T (1+B_p^2)$, respectively. In the Ohmic case the friction coefficient $\gamma$ is constant, while the diffusion coefficient $D_p$ is constant in the case of standard diffusion. An interesting relativistic example is photonic Janus particles with $m \equiv 0$. In this case Eq. (8) reduces to $\partial_t n_p = \partial_p (\gamma c n_p + D_p \partial_p n_p)$.

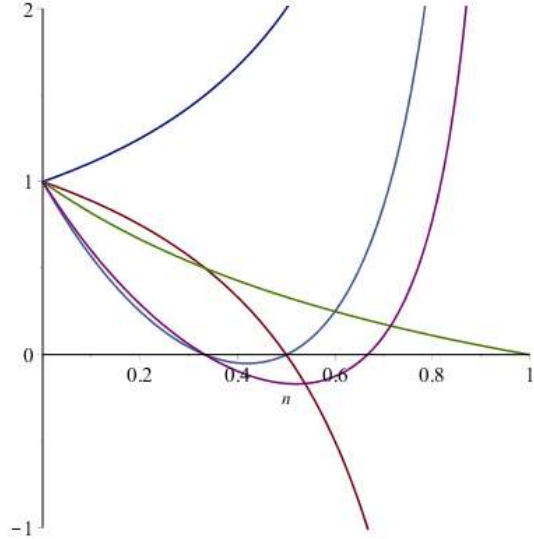

**Fig. 5** Plots of $D_p / \gamma k_B T$ for hexa-boson/fermion (light blue), penta-boson/half-fermion (purple), anti-fermion (brown), fermion (dark blue) and di-boson (green)

As is shown in the present paper, the entropic effects can lead to phase-transitions in the energy space. This is not surprising because the population of the quantum levels at a given temperature is governed by the entropy functional. The emerging entropic forces seem to be extremely strong, since they maintain the exclusion principle for fermions. The present analysis can be further generalized by employing the Tsallis entropy $S/k_B = (1/\alpha)\sum n_k (1-n_k^\alpha)$ [4], for instance, which contains the Boltzmann entropy as a particular case at $\alpha = 0$. Another interesting specific case of the Tsallis entropy is the linear entropy with $\alpha = 1$.

1. Walther A. and Müller, A.H.E. (2013) *Chem. Rev.* **113** 5194

2. Huang, K. (1963) *Statistical Mechanics*, Wiley, New York, p. 196

3. Jüttner, F. (1911) *Ann. Phys.* **339** 856

4. Tsallis, C. (1988) *J. Stat. Phys.* **52** 479